\documentclass[prd,amsfonts,amsmath,twocolumn,superscriptaddress,nofootinbib]{revtex4}

\usepackage{color}
\usepackage{graphicx} 
\usepackage{epstopdf}%

\usepackage{mathpazo}      

\newcommand {\be} {\begin{equation}} 
\newcommand {\ba}{\begin{eqnarray}} 
\newcommand {\ee} {\end{equation}} 
\newcommand{\ea} {\end{eqnarray}}

\renewcommand{\epsilon}{\varepsilon}

\begin{document}

\title{
Circular Dichroism of Twisted Photons in the Non-Chiral Atomic Matter}

\author{Andrei Afanasev}

\affiliation{Department of Physics,
The George Washington University, Washington, DC 20052, USA}

\author{Carl E.\ Carlson}
\affiliation{Department of Physics, The College of William and Mary in Virginia, Williamsburg, VA 23187, USA}

\author{Maria Solyanik$^1$}

\date{\today}

\begin{abstract}

We calculate the circular dichroism (CD) for absorption of the twisted photons, or optical vortices, by atoms, caused by atomic excitation into discrete energy levels.
The effects of photon spin on the rates and cross sections of atomic photo-excitation are considered. It is demonstrated that although for electric dipole transitions the atomic excitation rates  depend on the relative orientation of photon spin and orbital angular momentum (OAM), the resulting CD is zero. However, CD is nonzero for atomic transitions of higher multipolarity,  peaking in the optical vortex center, resulting in preferred absorption of the photons with their spins aligned with OAM. The effects remain large in a paraxial limit, where analytic expressions are provided. The predicted spin asymmetries are equivalent to OAM dichroism for the fixed photon spin.
\end{abstract}

\maketitle

\section{Introduction and Motivation}	\label{sec:intro}

Circular Dichroism (CD) is defined as differential absorption of left and right circularly polarized photons, and it is widely used in the analysis of materials that are non-symmetric under mirror transformation either through preferred chirality of the material structure or through magnetic phenomena \cite{Berova}.

Twisted photons, or optical vortices, carry OAM along their direction of propagation, and therefore they can be characterized by their own chirality (or topological charge); see, e.g., \cite{Padgett2015} for a recent review. Spin-dependence of OAM photon flux enters through a spin-orbit interaction \cite{BliokhAlonso10}.
Interactions of the twisted photons with non-chiral nano-structures were reported in Ref. \cite{Zambrana14, Zambrana16}, showing  significant CD, at 80 to 90 per cent level, for varied topological charges. While the observed large CD can be understood in principle by realizing that left and right circular polarization states are not mirror-symmetric for twisted photons, the observed effect still awaits theoretical explanation, likely in terms of surface plasmon dynamics \cite{Zambrana16}. Theoretical predictions of CD were previously made for twisted X-rays in metals \cite{PhysRevLett.98.157401}. 

In the present work we apply previously developed formalism for photo-excitation of atoms by the twisted photons
 \cite{Afanasev:2013kaa,Afanasev2014} and predict CD effects in the atomic matter. We take advantage is the observation 
 \cite{Afanasev:2013kaa,Scholz2014,Afanasev2016} that the transition amplitudes of atomic photo-excitation with twisted photons can be presented in a simple factorized form in terms of plane-wave photon amplitudes, making our predictions applicable to a variety of atomic targets. We present the results as a function of the distance between a given atom (or ion) to the center of the optical vortex, that we define as {\it an impact parameter} $b$. Precise measurements of $^{40}$Ca$^+$ ion excitations with the twisted light as a function of the impact parameter with sub-wavelength position resolution were performed recently 
Ref.\cite{Schmiegelow2016} using an ion trap. For the most recent review of twisted-light interactions with atoms, see \cite{Franke-Arnold2017} and references therein.

Sections II and III of this paper review the formalism for twisted photons and for calculating atomic photoexcitation with the twisted photons, respectively.  Sec. IV introduces CD and discusses photon spin effects in the the photoexcitation rates and cross sections, Sec.V describes the evolution of the twisted-photon polarization caused by absorption, Sec. VI provides the analytic expressions in paraxial limit,  and Sec. VII offers some closing comments.


\section{Definition of Twisted-Photon States}			\label{sec:one}


We define the twisted-photon states as non-paraxial Bessel beams according to~\cite{Jentschura:2010ap,Jentschura:2011ih}, that can be viewed as extensions of the nondiffractive Bessel modes described in~\cite{Durnin:1987,Durnin:1987zz}. More detail is given in~\cite{Afanasev:2013kaa}.   These states correspond to superposition of TE and TM Bessel modes introduced in Ref. \cite{Jauregui:2004}; see also Appendix of Ref.\cite{Afanasev2014} for detailed comparison.

A twisted photon state with symmetry axis passing through the origin, can be given as a superposition of plane waves and in Hilbert space can be written as, 
\begin{align}
\label{eq:twisteddefinition}
| \kappa m_\gamma k_z \Lambda \rangle 
&= \sqrt{\frac{\kappa}{2\pi}} \  \int \frac{d\phi_k}{2\pi} (-i)^{m_\gamma} e^{im_\gamma\phi_k}  \,
	|\vec k, \Lambda\rangle		\,.
\end{align}
The component states on the right are plane-wave states, all with the same longitudinal momentum $k_z$, the same transverse momentum magnitude $\kappa = |\vec k_\perp|$, and the same plane wave helicity $\Lambda$ (in the directions $\vec k$).  The angle  $\phi_k$  is the azimuthal angle of vector $\vec k$, and with the phasing shown, $m_\gamma$ is the total angular momentum in the $z$ direction. We also define a pitch angle   $\theta_k = \arctan (\kappa/k_z)$, and $\omega = | \vec k |$.  The pitch angle $\theta_k$ was first introduced in the definition of Bessel beams in Ref.\cite{Durnin:1987,Durnin:1987zz}, and it is related to Berry phase of the photon as \cite{BliokhAlonso10}: $\Phi_B=2\pi(1-\cos\theta_k)$. The phase singularity of this beam is located on the beam symmetry axis.

The electromagnetic potential of the twisted photon in coordinate space is 
\begin{align}
\label{eq:twistedwave}
\mathcal A^\mu_{\kappa m_\gamma k_z \Lambda}(t,\vec r)
&= \sqrt{\frac{\kappa}{2\pi}} \, e^{-i\omega t} \nonumber\\
&\times	\int \frac{d\phi_k}{2\pi} (-i)^{m_\gamma} e^{im_\gamma\phi_k}  \,
	\epsilon^\mu_{\vec k,\Lambda} e^{i \vec k {\cdot} \vec r}	.
\end{align}
The polarization vectors are 
\be
\label{eq:epsilonexpand}
\epsilon^\mu_{\vec k \Lambda} \!\! = \!
	e^{-i\Lambda\phi_k} \! \cos^2\frac{\theta_k}{2} \eta^\mu_\Lambda
	+ e^{i\Lambda\phi_k} \! \sin^2\frac{\theta_k}{2} \eta^\mu_{-\Lambda}
	+ \frac{\Lambda}{\sqrt{2}} \sin\theta_k  \eta^\mu_0
\ee
with $4$-dimensional unit vectors,
\be
\eta^\mu_{\pm 1} = \frac{1}{\sqrt{2}}  \left( 0,\mp 1,-i,0 \right)	\,,
\quad \eta^\mu_0 =  \left( 0,0,0,1 \right)	\,.
\ee
The electromagnetic potential then has a form
\begin{align}
\label{eq:twistedwf}
\mathcal A^\mu_{\kappa m_\gamma k_z \Lambda}(x) &= e^{-i(\omega t - k_z z)}	
\sqrt{\frac{\kappa}{2\pi}}  \, \nonumber\\ &	\Bigg\{
	\frac{\Lambda}{\sqrt{2}} e^{im_\gamma\phi_\rho} \sin\theta_k
	J_{m_\gamma}(\kappa\rho) \, \eta^\mu_0			\nonumber\\[1ex]
& \quad + i^{-\Lambda} e^{i(m_\gamma-\Lambda)\phi_\rho}  \cos^2\frac{\theta_k}{2} 
	J_{m_\gamma-\Lambda}(\kappa\rho) \, \eta^\mu_\Lambda	\nonumber\\[1ex]
& \quad + i^{\Lambda}  e^{i(m_\gamma+\Lambda)\phi_\rho}  \sin^2\frac{\theta_k}{2} 
	J_{m_\gamma+\Lambda}(\kappa\rho) \, \eta^\mu_{-\Lambda}
	\Bigg\}	\,.
\end{align}

The energy flux is given by
\begin{align}
\label{eq:flux}
f(\rho)  &= \cos(\theta_k) (|E|^2+|B|^2)/4= 
 \cos(\theta_k) \frac{\kappa\omega^2 }{2\pi} \nonumber\\ &
 \Bigg\{ \cos^4\frac{\theta_k}{2} J^2_{m_\gamma-\Lambda}(\kappa\rho) +\sin^4\frac{\theta_k}{2} J^2_{m_\gamma+
\Lambda}(\kappa\rho) \\ &
+\frac{\sin^2\theta_k}{2}J^2_{m_\gamma}(\kappa\rho) \Bigg\} . \nonumber 
\end{align}
The use of the above canonical-momentum expression is essential, since Poynting vector alone does not represent the full energy flux of a twisted photon beam \cite{BliokhAlonso10,Bliokh2015}.


\section{Plane-wave factorization for atomic photoexcitation with twisted photons}			


Here we briefly review the formalism of atomic photoexcitation by the twisted photons worked out previously \cite{Afanasev:2013kaa,Afanasev2014,Scholz2014, Afanasev2016}, leading to plane-wave factorization property of the twisted-photon absorption.

\begin{figure}[t]
\begin{center}
\includegraphics{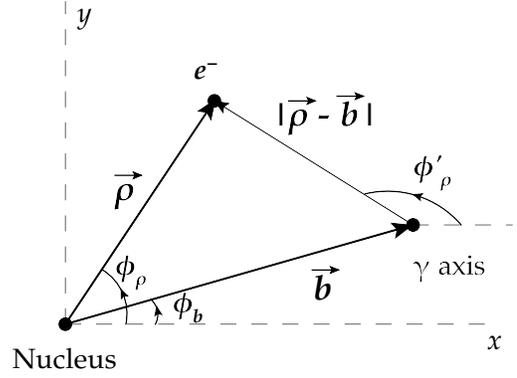}
\caption{Relative positions of atomic state and photon axis, as projected onto the $x$-$y$ plane, with the origin at the nucleus of the atom.}
\label{fig:Atom}
\end{center}
\end{figure}

Consider the excitation by a twisted photon of an atom. The photon's wave front travels in the $z$-direction and the axis of the twisted photon is displaced from the nucleus of the atomic target by some distance in the $x$-$y$ plane which we will call an impact parameter $\vec b$, Fig.\ref{fig:Atom}. The transition matrix element is
\begin{align}
\label{eq:Smatrix}
S_{fi} &= -i \int dt  \langle n_f l_f m_f | H_1 
	| n_i l_i m_i; \kappa m k_z \Lambda \rangle				\,,
\end{align}
where the non-relativistic interaction Hamiltonian is given by
\be
\label{eq:hamiltonian}
H_1 = - \frac{e}{m_e} \vec A \cdot \vec p \,,
\ee
and we use standard notation $(n,l,m)$ for the principal, orbital and magnetic quantum numbers of initial and final states of an atom.

It can be shown that  Ref.\cite{Afanasev:2013kaa,Scholz2014,Afanasev2016} that for atomic excitation from the ground state ($l_i=m_i=0$) the above amplitude from Eq.(\ref{eq:Smatrix}) is proportional to the plane-wave amplitude $\mathcal M^{\rm (pw)}$ times Wigner $d$-functions that only depend on the pitch angle $\theta_k$ and Bessel functions that define the amplitude dependence on the impact parameter $b$. Here, the Bessel factor arises due to the azimuthal phase dependence of the twisted-photon and the excited atomic state, and Wigner $d$-functions is a result of a tilted quantization axis (by an angle $\theta_k$) with respect to the direction of beam propagation:
\begin{align}
\label{eq:factorized}
&|{\cal M}_{n_f l_f m_f \Lambda}(b)|   =  \nonumber \\  &\left |\sqrt{\frac{\kappa}{2\pi}}J_{m_f-m_\gamma}(\kappa b)d^{l_f}_{m_f\Lambda}(\theta_k) \mathcal M^{\rm (pw)}_{n_f l_f \Lambda \Lambda}(\theta_k=0)\right |
\end{align}

The factorized form of the transition amplitude facilitates comparison of twisted-photon vs plane-wave absorption by atoms. It contains the details of atomic structure in a common-factor plane-wave amplitude $\cal M^{\rm (pw)}$, while the novel features arising from the phase and spatial structure of the twisted light are contained in Wigner and Bessel functions that enter  independently of the specific details of atomic wave functions.  When deriving the factorization property Eq.(\ref{eq:factorized}), we used the first-order Born matrix element that assumes the interaction proceeds in the linear regime, justifying representation of S-matrix as a linear superposition of plane-wave matrix elements. To further prove total angular momentum conservation under photo-absorption, we previously assumed that the atom is much smaller than the wavelength of absorbed light \cite{Afanasev2014}, but this assumption is not needed to derive Eq.(\ref{eq:factorized}). Therefore as long as the linear interaction regime holds, the above formula is applicable to twisted-photon excitation of arbitrary quantum systems, such as atoms, molecules, ions, atomic nuclei, excitons or quantum dots. For example taking the limit $\theta_k\to\pi/2$ in Eq.(\ref{eq:factorized}), we recover a similar factorization property recently derived for the absorption of polariton vortices, $c.f.$ Eq.(2) of  \cite{Machado2016}.

\section{Spin-Dependence and Circular Dichroism of Twisted-Photon Absorption}

The twisted-photon flux depends on the sign of $\Lambda$ defining the handedness of plane-wave photons that form a given Bessel beam, and this dependence was discussed previously by Bliokh and collaborators in the context of spin-orbit interaction of light \cite{BliokhAlonso10,BliokhNature15}.

The photo-excitation cross section of is given by 
\begin{equation}
\label{eq:xsec}
\sigma^{(m_\gamma)}_{n_f l_f \Lambda} = 2 \pi \delta(E_f -E_i-\omega_\gamma)
 \frac{ \sum_{m_f = - l_f}^{m_f=l_f}|\mathcal M^{(m_\gamma)}_{n_f l_f m_f \Lambda}(b)|^2 }{f} \,.
\end{equation}
where summation over final spins and averaging over initial spins is implied and the photon flux $f$ may be either unintegrated $f(b)$ (given by Eq.(\ref{eq:flux})) or integrated over the transverse beam profile. For the excitation rates $\Gamma$ one removes the flux $f$ from the above expression, $i.e.$
\begin{equation}
\label{eq:rate}
\Gamma^{(m_\gamma)}_{n_f l_f \Lambda}=f\cdot \sigma^{(m_\gamma)}_{n_f l_f  \Lambda} .
\end{equation}

Circular dichroism is defined as a differential absorption probability for left- vs right-circularly polarized light. It can be observed either by measurements of polarization dependence of light transmission through matter or by observing ellipticity acquired by a linearly polarized light beam as a result of the difference of absorption in $\Lambda=+1$ and $\Lambda=-1$ states.  

Let us compare the rates and the cross sections of photo-absorption of the twisted photons with opposite signs of circular polarization $\Lambda$, while keeping (paraxial-limit) OAM unchanged. It leads to a definition
\begin{equation}
\label{eq:CD}
CD^{(\overline m_\gamma,l_f)}=\frac{\sigma^{(\overline m_\gamma+1)}_{n_f l_f; \Lambda=1}-\sigma^{(\overline m_\gamma-1)}_{n_f l_f; \Lambda=-1}}
{\sigma^{(\overline m_\gamma +1)}_{n_f l_f; \Lambda=1}+\sigma^{(\overline m_\gamma-1)}_{n_f l_f; \Lambda=-1}},
\end{equation}
where $\overline m_\gamma$ defines twisted photon's topological charge that corresponds to its OAM projection in a paraxial limit.

Using Equations (\ref{eq:flux},\ref{eq:factorized}), we first consider electric dipole $E1$-transitions for which OAM of the atomic electron changes by one unit, and find CD to be identically zero for any twisted light beam of a given $\overline m_\gamma$. We also find from the same equations that CD is zero for beams with $\overline m_\gamma$=0 and any $l_f$, which is expected from parity considerations. 
\begin{equation}
CD^{(\overline m_\gamma,l_f=1)}=CD^{(\overline m_\gamma=0,l_f)}=0.
\end{equation}

The reason behind this null result for $E1$-transitions is that excitation rates $\Gamma$ for these transitions are proportional to the photon flux $f(b)$, and even though they separately depend on the sign of $\Lambda$, this dependence cancels in the cross section, yielding zero CD for isotropic atomic targets.

\begin{figure}[h]
\begin{center}
\includegraphics[width = 84 mm]{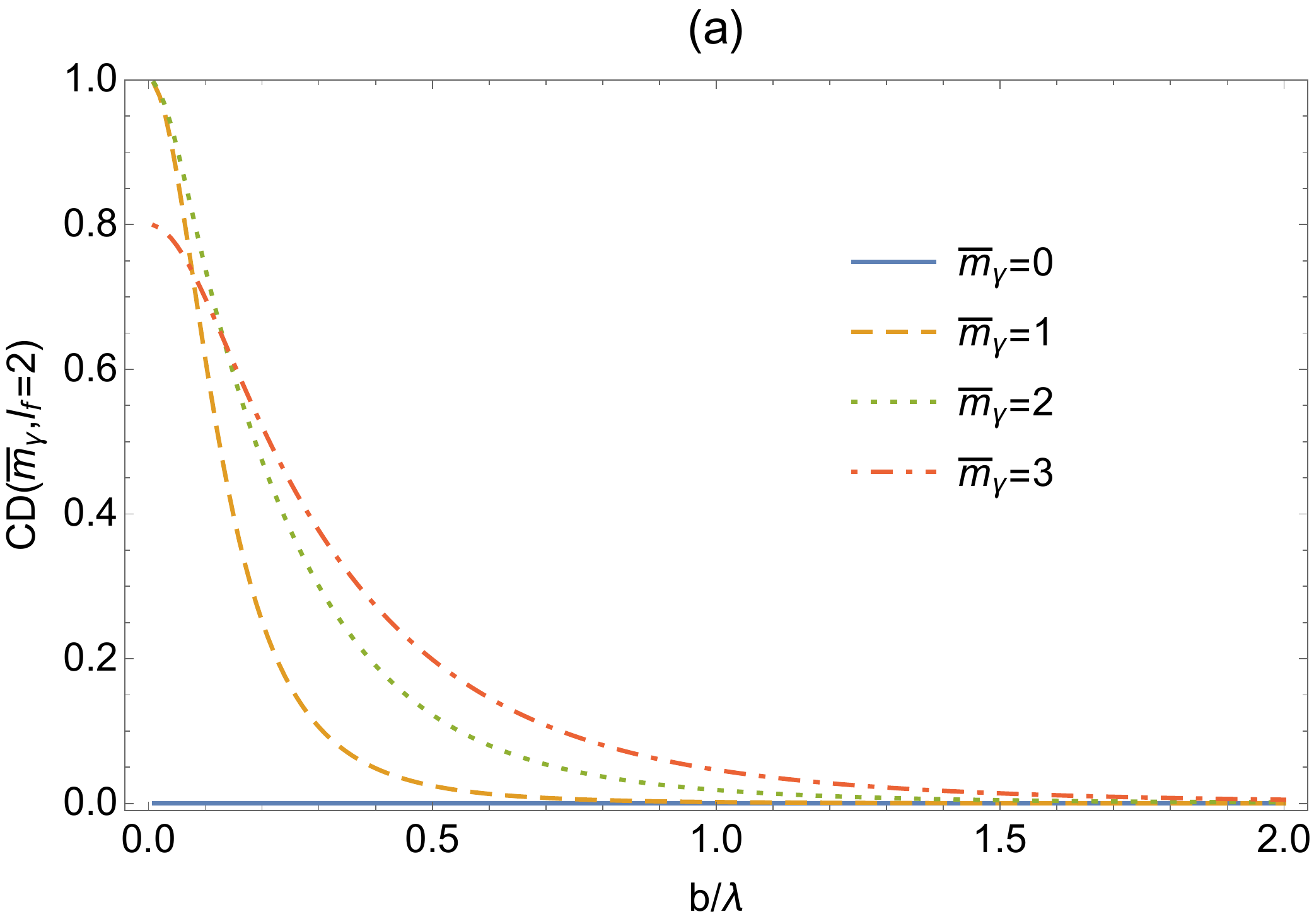}
\includegraphics[width = 84 mm]{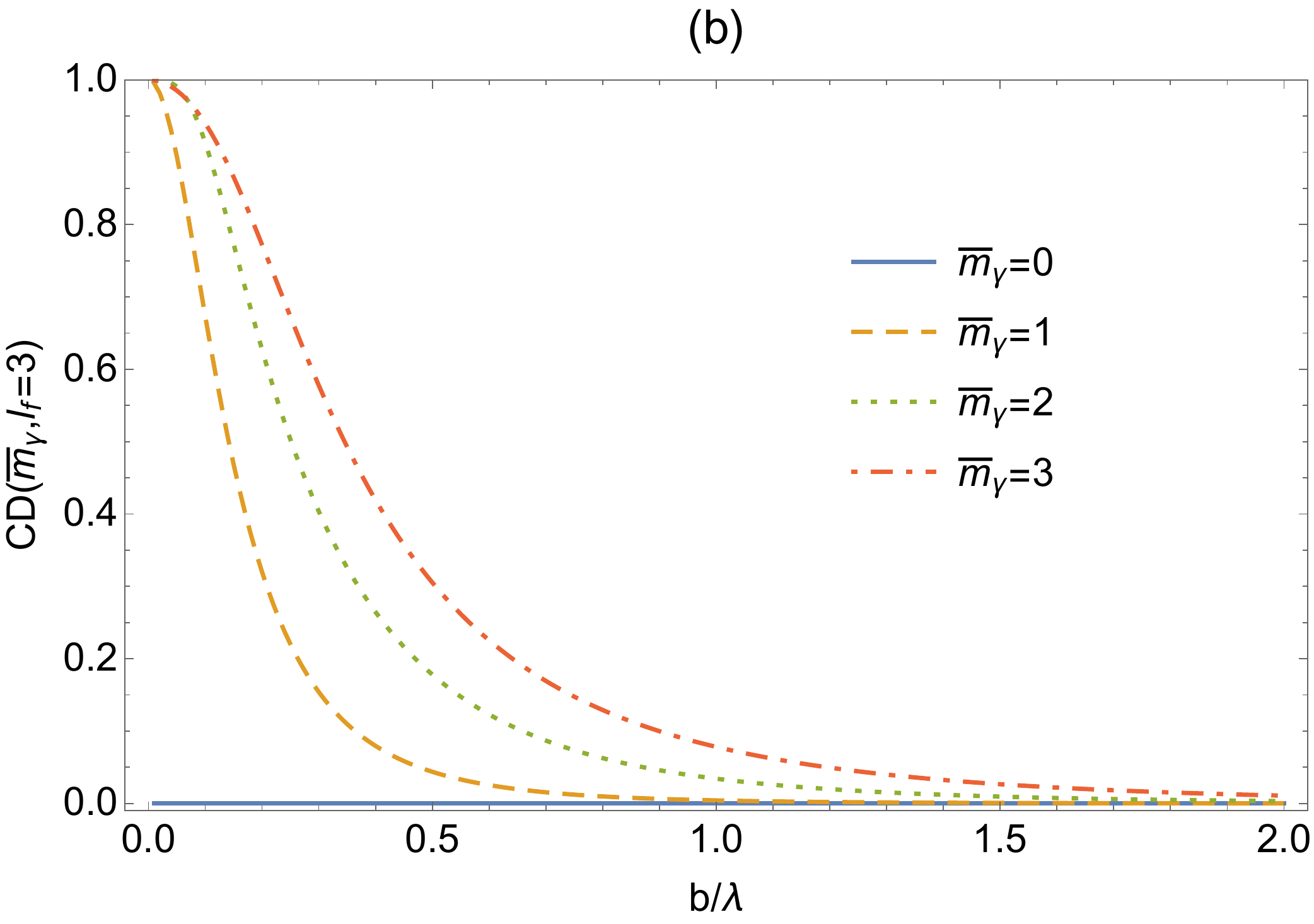}
\caption{Circular dichroism (CD) as a function of impact parameter $b$ (in units of photon wavelength $\lambda$) for different values of $m_\gamma$ for excitation of the atomic states with $l_f=2$ (a) and $l_f=3$ (b); the angle $\theta_k$=0.1 rad. The curve styles denote the average photon's angular momentum projection: $\overline m_\gamma=0$ is the blue solid curve, $\overline m_\gamma=1$ is orange and dashed, $\overline m_\gamma=2$ is green and dotted, $\overline m_\gamma=3$ is red and dot-dashed. CD is zero for the beams with no OAM ($\overline m_\gamma$=0). }
\label{fig:CD}
\end{center}
\end{figure}

\begin{figure}[h]
\begin{center}
\includegraphics[width = 84 mm]{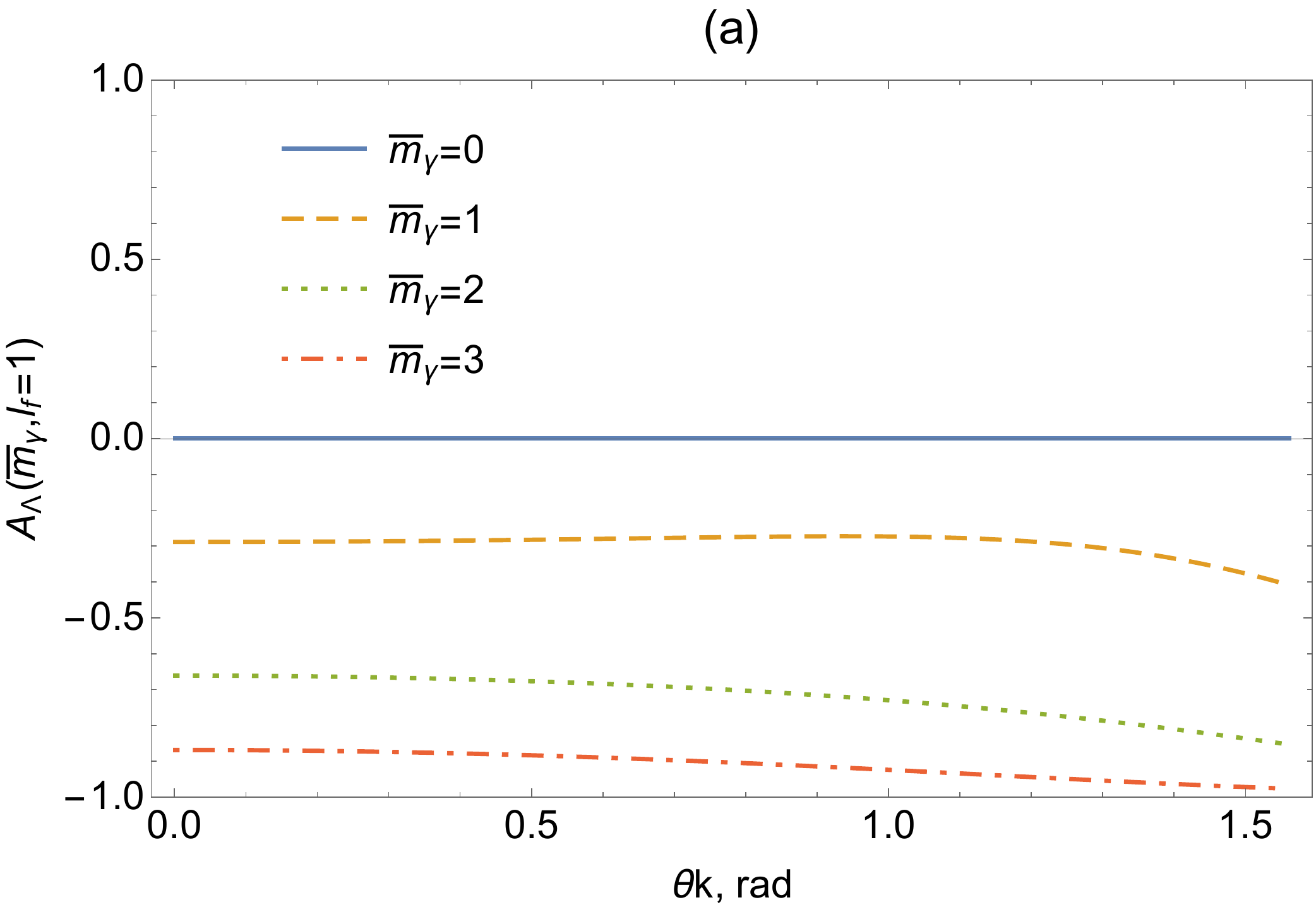}
\includegraphics[width = 84 mm]{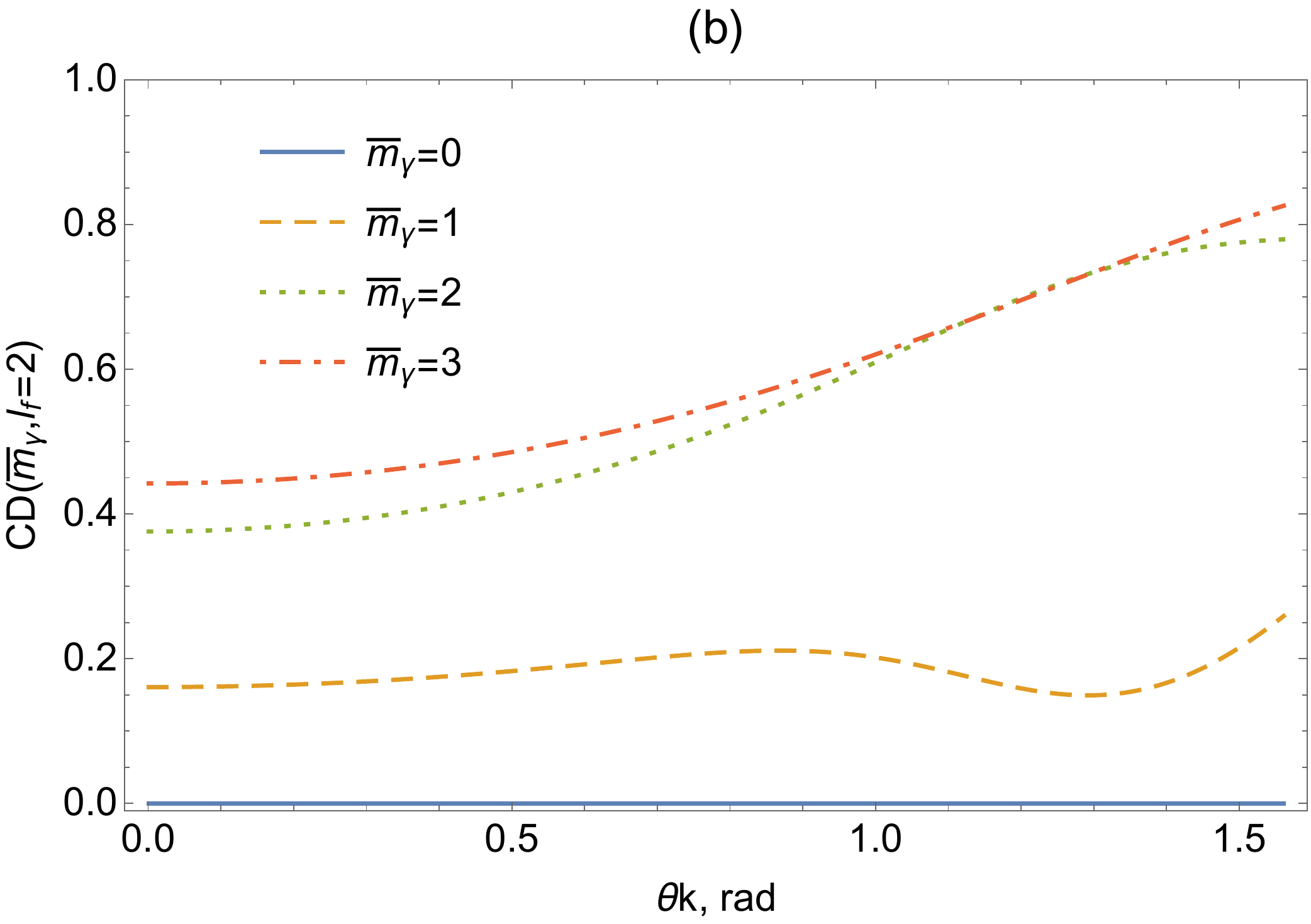}
\caption{CD as a function of a pitch angle $\theta_k$ for a fixed impact parameter $b/\lambda=0.25$, for (a) electric dipole $E1$ transitions ($l_f=1$), which coincides with spin asymmetry of the photon flux Eq.(\ref{eq:flux}) and (b) electric quadrupole $E2$ transitions ($l_f=2$). Different curves correspond to different topological charges $\overline m_\gamma$, the notation is as in Fig.\ref{fig:CD}.}
\label{fig:CDang}
\end{center}
\end{figure}

The calculation results for CD are shown in Fig.\ref{fig:CD} for electric quadrupole $\Delta l=2$  (a) and electric octupole $\Delta l=3$ transitions (b) for a fixed pitch angle $\theta_k$=0.1 rad. The calculations show large and positive values of CD near the phase singularity of the optical vortex $b \to 0$, and CD is seen to die off at the atom's positions of about one photon wavelength and larger. It means that a twisted photon whose spin is {\it aligned}  with its OAM have a higher relative probability to be absorbed by the atom. Further analyzing the dependence on the pitch angle $\theta_k$ we find that CD becomes independent of this parameter in a broad range of moderately small angles, and remains the same in the paraxial limit $\theta_k \to 0$, see Fig.\ref{fig:CDang}. If the experiment does not resolve the atom's position $b$, we would have to integrate over all impact parameters, which would result in  zero CD due to the fact that total excitation cross sections would not depend on a particular value of $m_\gamma$ or $\Lambda$, see Ref.\cite{Afanasev:2013kaa}.

In experiments that directly measure the atomic excitation rates as a function of the impact parameter $b$, the results are presented as rates (or Rabi frequencies as in Ref.\cite{Schmiegelow2016}) normalized to the total laser-beam power within an aperture that is much wider that the wavelength of light. In this case a relevant observable would be $\Lambda$-dependence of the {\it photo-excitation rate}  that was analyzed theoretically in Ref.\cite{Afanasev:2013kaa} (for the case of leading $E1$ transitions).

We define photon-spin asymmetry of the photo-excitation rate similarly to CD, 
\begin{equation}
\label{eq:SpinAsym}
A_\Lambda^{(\overline m_\gamma,l_f)}=\frac{\Gamma^{(\overline m_\gamma+1)}_{n_f l_f;\Lambda=1}-\Gamma^{(\overline m_\gamma-1)}_{n_f l_f; \Lambda=-1}}
{\Gamma^{(\overline m_\gamma +1)}_{n_f l_f; \Lambda=1}+\Gamma^{(\overline m_\gamma-1)}_{n_f l_f; \Lambda=-1}}.
\end{equation}

\begin{figure}[t]
\begin{center}
\includegraphics[width = 84 mm]{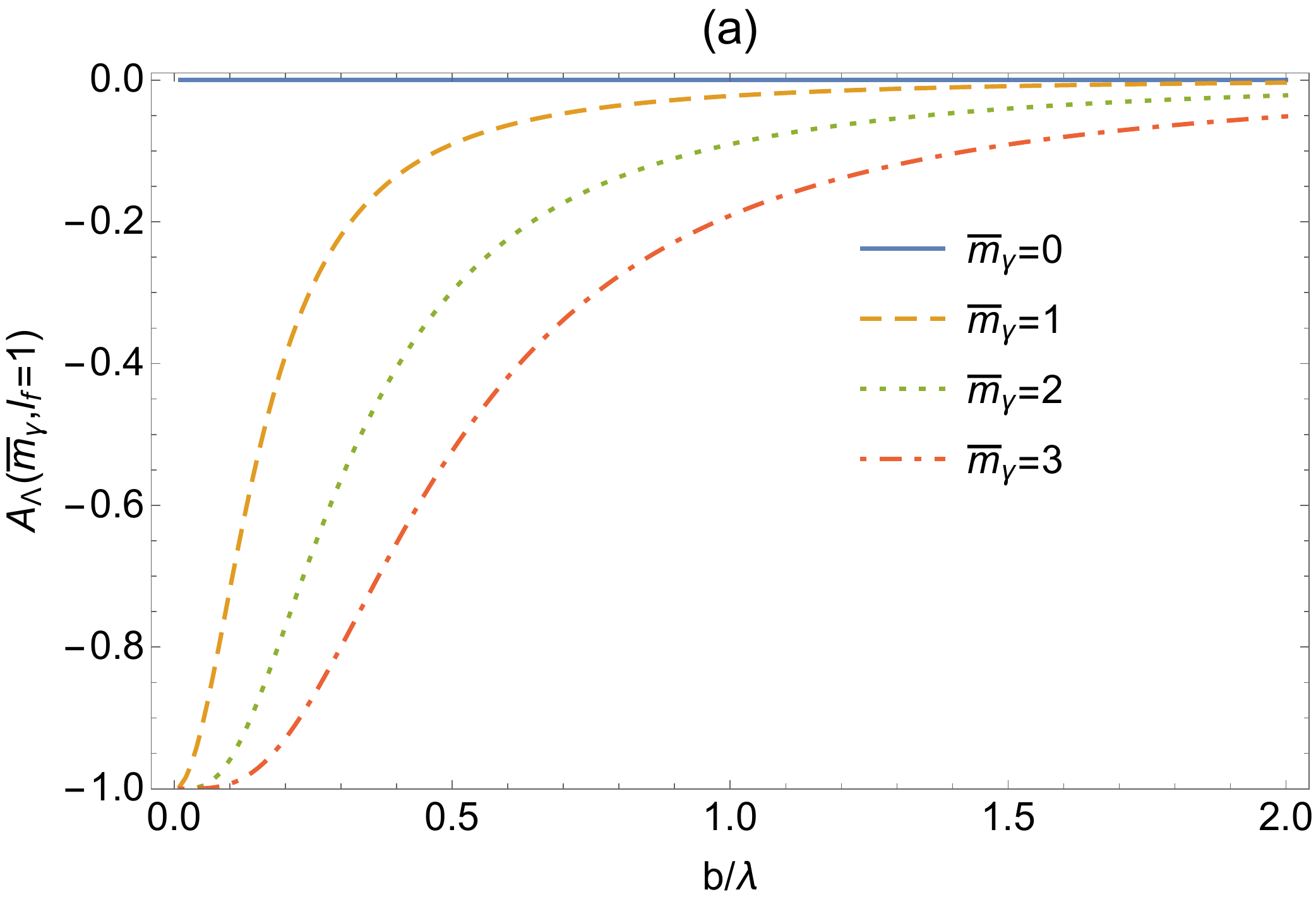}
\includegraphics[width = 84 mm]{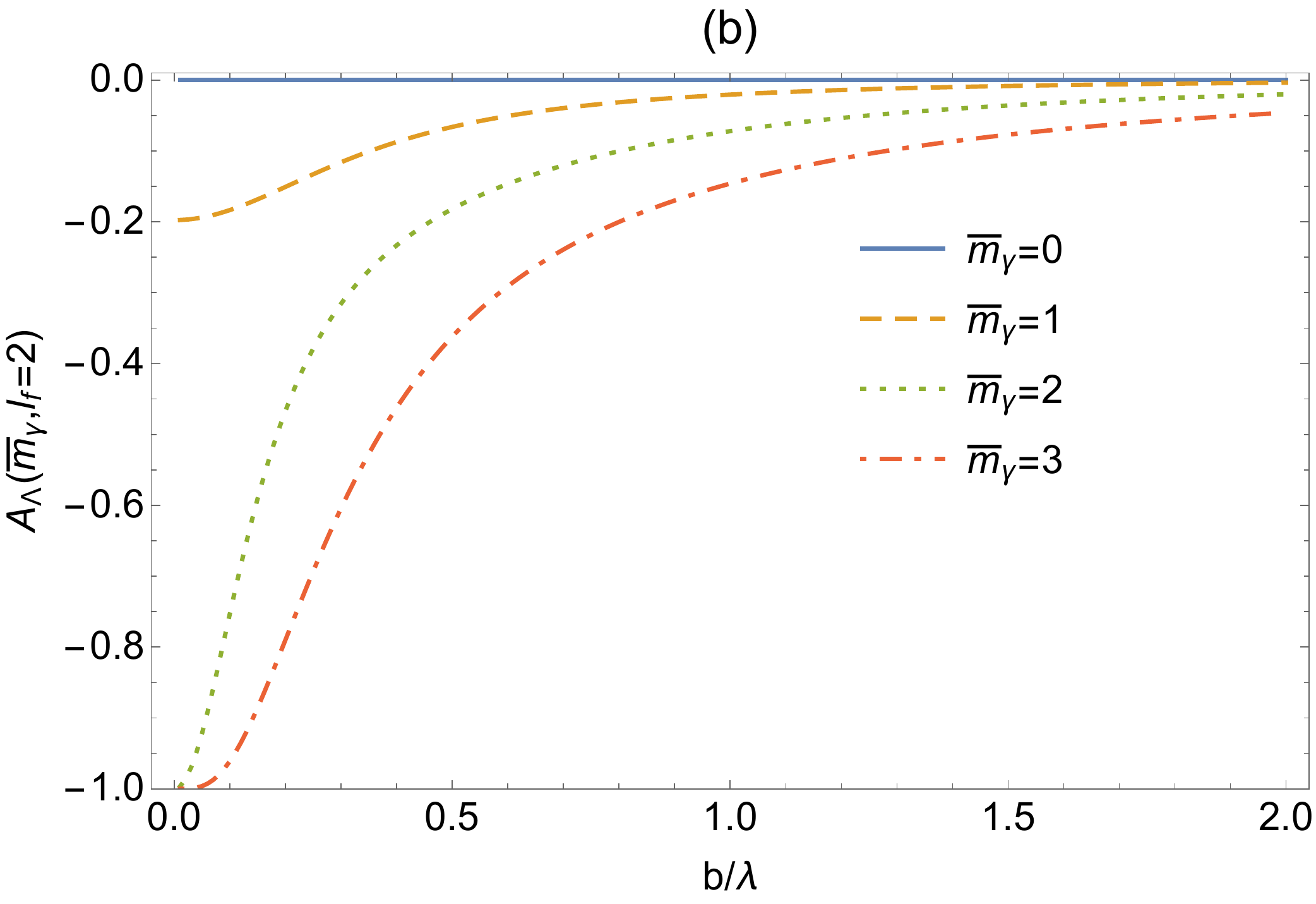}
\includegraphics[width = 84 mm]{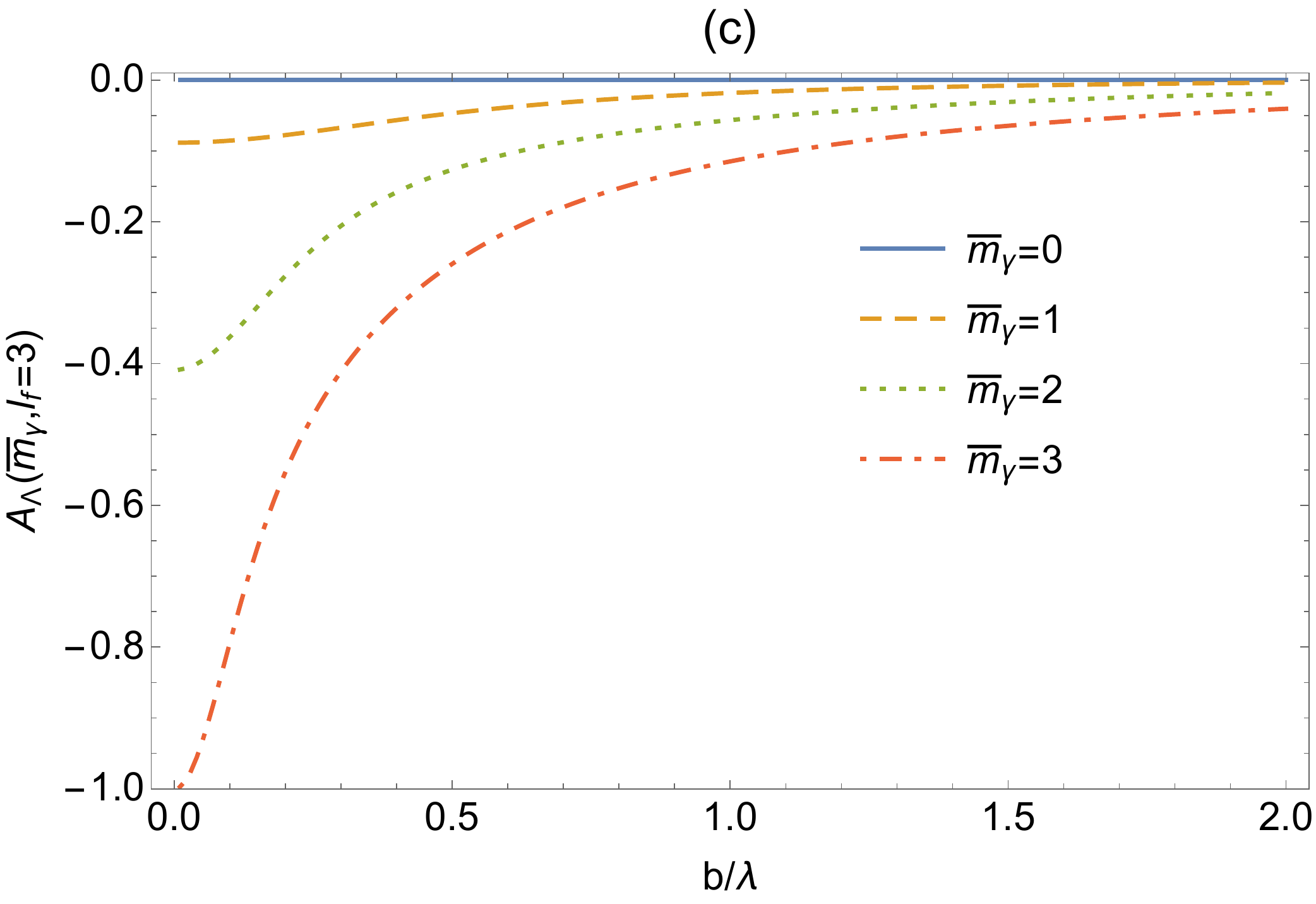}
\caption{Spin asymmetry of the photoexcitation rates for electric dipole E1 (a) quadrupole $E2$ (b) and octupole E3 (c) transitions. Different curves correspond to different topological charges $\overline m_\gamma$, the notation is as in Fig.\ref{fig:CD}, and the angle $\theta_k$=0.1 rad.}
\label{fig:rateAs}
\end{center}
\end{figure}

The results are presented in Fig.\ref{fig:rateAs} for the atomic transitions into the states of $l_f=1,2,$ and 3 caused by the photons with  topological charges $\overline m_\gamma$=0 to 3.
We can see that the spin asymmetry of rates behaves differently from CD: The asymmetry is negative within the distance of about one wavelength near the optical vortex center and, with a few exceptions, reaches a value of -1 at the center. It means that the transitions at the vortex center are mainly caused by the twisted photons whose spin and OAM are {\it anti-aligned}.  An apparent difference from positive-sign CD is due to the fact that there is a relatively denser flux of the {\it anti-aligned} twisted photons at the vortex center for the same overall beam power.

 If the atom's position is not resolved, we would have to integrate over the position, resulting into zero spin asymmetry $A_\Lambda$, similarly to the above result for CD. It implies that {\it observation of spin-asymmetric absorption of twisted light by atomic matter requires localization of the target atoms within about light's wavelength}. It can be achieved, for example, by using nano-sized apertures, well-localized ions in Paul traps, or mesoscopic targets.


\section{Evolution of the Twisted-Light Polarization under Propagation in Matter}

Above predictions of nonzero CD for the twisted light being absorbed by atoms would lead to the evolution of twisted-photon polarization states.
Indeed, let us represent an arbitrary polarization state as a superposition of $\Lambda=1$ and $\Lambda=-1$ spin states, or left- and right- circularly polarized states, for a given topological charge:
\begin{equation}
\label{eq:stateevol}
|\kappa \overline m_\gamma k_z>=c_{-} |\kappa \overline m_\gamma k_z \Lambda=-1>+c_{+} |\kappa \overline m_\gamma k_z \Lambda=1>
\end{equation}
where $c_{\pm}$ are complex coefficients. Their dependence on the propagation distance $z$ is controlled by the attenuation coefficients $\mu_{\pm}$, that in turn are proportional to the photoabsorption cross sections $\mu_{\pm}= \sigma_\pm \cdot n$, where $n$ is a number of atoms per unit volume, with the expressions for $\sigma_\pm$ coming from Eq.(\ref{eq:xsec}). Since we are interested in comparison with the plane-wave propagation, we can express the attenuation coefficients in terms of their plane-wave values and the cross section ratios $r_\pm^{tw}(b)$ introduced in \cite{Afanasev:2013kaa,Afanasev2016}:
\begin{equation}
\mu_\pm=\mu^{pw}\frac{\sigma}{\sigma^{pw}}=\mu^{pw}r_{\pm}^{tw}(b),
\end{equation}
where the plane-wave attenuation coefficient $\mu^{pw}$ is independent of $\Lambda$ for the isotropic atomic matter, while the factor $r_\pm^{tw}(b)$ depends on  $\Lambda$ and on the impact parameter $b$. Then
\begin{equation}
c_{\pm}(z)=c_{\pm}(0) e^{-\mu_\pm z/2}=c_{\pm}(0) e^{-\mu^{pw}z r_\pm^{tw}(b)/2}.
\end{equation}

For the case of electric dipole transitions $l_f$=1, it follows from Eqs.(\ref{eq:flux},\ref{eq:factorized}) that $r_\pm^{tw}(b)=1/\cos\theta_k$, independently of $\Lambda$ and $b$ \cite{Afanasev:2013kaa,Scholz2014, Afanasev2016}. Therefore, the coefficients $c_{\pm}(z)$ have the same $z$-dependence resulting in no evolution of the twisted-photon polarization due to atomic absorption via electric-dipole transitions. However for higher-multipole transitions into the states $l_f>1$ the factors $r_\pm^{tw}(b)$ depend on the photon spin projection due to CD as defined in Eq.(\ref{eq:CD}) (since the plane-wave cross section is a constant and it cancels in the ratio). 

\begin{figure}[t]
\begin{center}
\includegraphics[width = 84 mm]{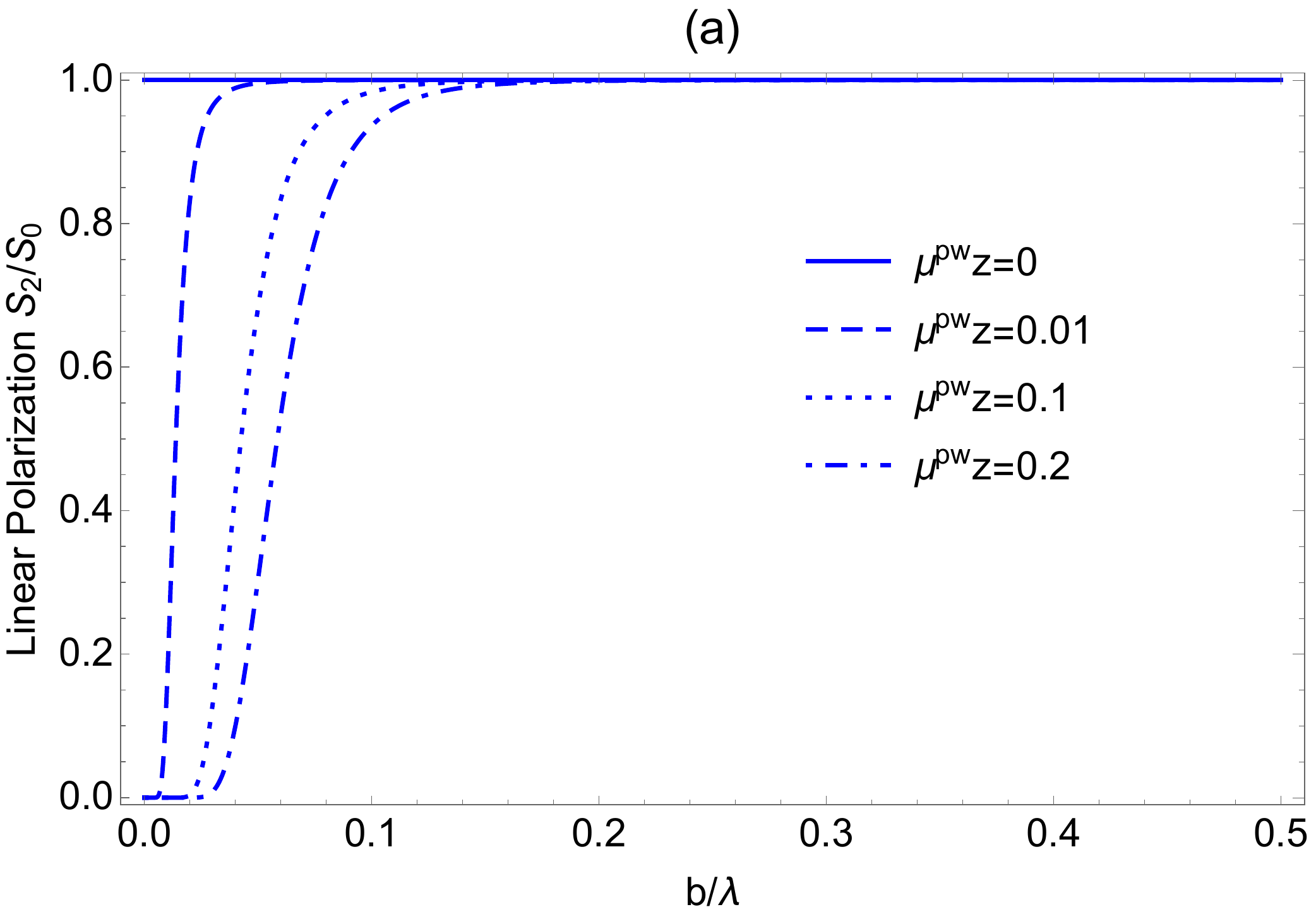}
\includegraphics[width = 84 mm]{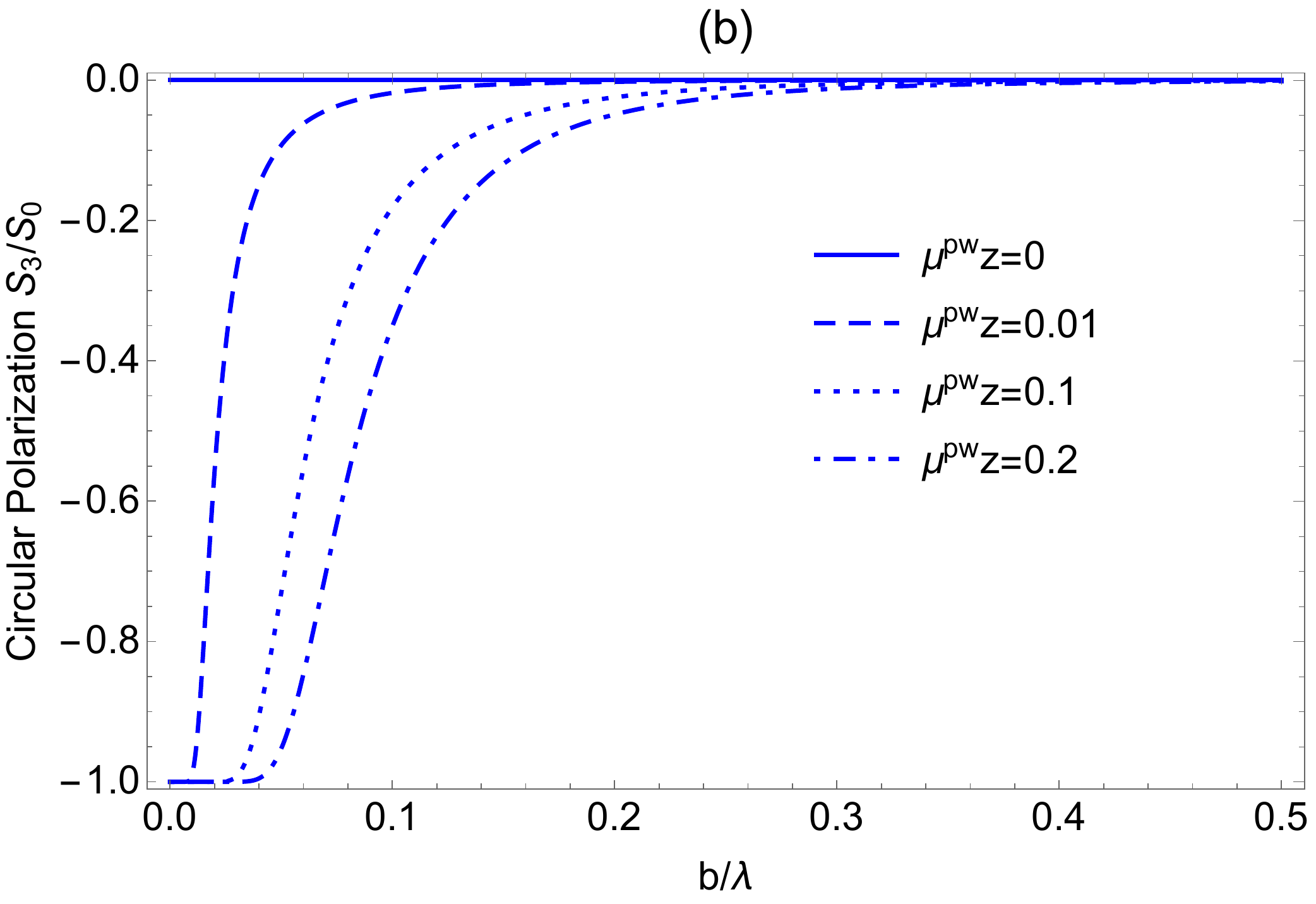}
\caption{Evolution of linear polarization $S_2/S_0$ (a) and circular polarization $S_3/S_0$ (b) of an optical vortex due to absorption in the atomic matter as a function of impact parameter $b$ for different values of propagation distance $z$=0 (solid blue), z=0.01 (dashed), z=0.1 (dotted) and z=0.2 (dot-dashed line), where $z$ is in units of plane-wave attenuation length $1/\mu^{pw}$. A topological charge is $\overline m_\gamma=1$ and the angle is $\theta_k$=0.1 rad. }
\label{fig:PolEvol}
\end{center}
\end{figure}

For example, let us consider a superposition of $\Lambda=1$ and $\Lambda=-1$ spin states for the same topological charge $\overline m_\gamma=1$, and assume the coefficients to be real and initially equal: $c_-|_{z=0}=c_+|_{z=0}$, where the field potential with a given $\Lambda$ is defined by Eq.(\ref{eq:twistedwf}). It results in a state with almost 100\% linear polarization of transverse fields in the central region of the vortex (except for the small region near the node of Bessel function $J_{\overline m_\gamma}$). However, even for small values of propagation distance $z$, the optical vortex develops 100\% circular polarization in the vortex center ($C$-point), and this region broadens as the beam passes further through the atomic matter. This prediction is shown in Fig.\ref{fig:PolEvol}, where we used standard definitions for Stokes parameters $S_{0-3}$ \cite{FieldGuidePolarization}. Development of $C$-type polarization singularity at the vortex center as a result of beam propagation can be observed in a dedicated experiment.

It should be noted that here we only considered the effects from photon absorption that result in CD. An additional effect would be forward scattering of the twisted photons, see, e.g. \cite{Davis13}. Corresponding spin dependence would result in circular birefringence showing in rotation of polarization plane of the linearly-polarized twisted light due to spin dependence of refractive index.

\section{Spin-Dependence and Circular Dichroism in Paraxial Limit}

Numerical calculations of spin-dependent observables for the twisted light with a fixed topological charge $\overline m_\gamma$ reveal smooth transition to the paraxial limit $\theta_k\to 0$. Actually, as seen in Fig.\ref{fig:CDang}, there is little dependence on this angle for $\theta_k \leq $ 0.25 rad. This observation prompts us to apply small-angle Taylor expansion for the expressions of CD and $A_\Lambda$ using explicit formulae Eqs.(\ref{eq:flux},\ref{eq:factorized}). Noticing that the argument of Bessel functions is 
$\kappa b=k \sin \theta_k b \approx k b \theta_k$ and defining $k b \equiv x$, we consider $\Lambda$-dependence of the photon flux Eq.(\ref{eq:flux}), which is the same as the rate of $l_f=1$ dipole excitation.

For the case of topological charge $\overline m_\gamma=1$ we have after Taylor-expanding Bessel functions for small values of their arguments:
\begin{align}
\label{eq:intTaylor}
& m_\gamma=2, \  \Lambda=1,  \ \ f_{(\Lambda=1)}\propto \frac{x^2}{4} \theta_k^2,  \nonumber\\
& m_\gamma=0, \  \Lambda=-1,\ \ f_{(\Lambda=-1)}\propto \frac{x^2}{4} \theta_k^2+\frac{\theta_k^2}{2}, \\ \nonumber
\end{align}
where the $x$-independent term in the last row comes from $J_{m_\gamma}$ term in the flux Eq.(\ref{eq:flux}), the term being indicative of spin-orbit interaction \cite{BliokhAlonso10}.

The dependence on $\theta_k $ cancels in the expression for the asymmetry, yielding
\begin{equation}
\lim_ {\theta_k\to 0}A_\Lambda^{(\overline m_\gamma=1,l_f=1)}=\frac{-1}{1+x^2},
\end{equation}
where the $x-independent$ spin-orbit term of Eq.(\ref{eq:intTaylor}) is a cause for a nonzero asymmetry. In view the relevant discussion in the literature on the spin-orbit interaction of light \cite{BliokhAlonso10, BliokhNature15}, we can attribute nonzero spin effects to this interaction.

Similarly, we can obtain the expression for  $A_\Lambda$ for any topological charge:
\begin{equation}
\lim_ {\theta_k\to 0}A_\Lambda^{(\overline m_\gamma,l_f=1)}=\frac{-1}{1+\frac{2 x^4/\overline m_\gamma^2}{(\overline m_\gamma-1)^2+2x^2}},
\end{equation}
where in general all three terms of Eq.(\ref{eq:flux}) appear to be of the same order in $\theta_k$ for the anti-aligned spin and OAM.

This expression closely matches rate asymmetries in Fig.\ref{fig:rateAs}(a) for moderately small pitch angles $\theta_k\leq 0.25$. It is remarkable that while the beam waist size strongly depends on the angle $\theta_k$ (which in turn relates to Berry phase \cite{BliokhAlonso10}), the spin asymmetry in the paraxial limit depends only on the topological charge $\overline m_\gamma$.

We can use the same approach to determine the spin asymmetries of photoexcitation rates for higher transition multipolarities, that in addition requires small-angle expansion of Wigner $d$-functions in Eq.(\ref{eq:factorized}). For example, for the quadrupole transitions caused by $\overline m_\gamma$=1 beam, we identify the terms that are leading-order in small $\theta_k$-expansion:
\begin{align}
& m_\gamma=2, \  \Lambda=1,  \nonumber \\
&{\cal M}{(m_f=2)}\propto J_0(\kappa b)\cdot d^{(2)}_{21}(\theta_k)\approx \theta_k, \\ \nonumber
&{\cal M}{(m_f=1)}\propto J_1(\kappa b)\cdot d^{(2)}_{11}(\theta_k)\approx k \theta_k/2, \\
& m_\gamma=0, \  \Lambda=-1,  \nonumber \\
&{\cal M}{(m_f=0)}\propto J_0(\kappa b)\cdot d^{(2)}_{0-1}(\theta_k)\approx \sqrt{\frac{3}{2}}\theta_k, \\ \nonumber
&{\cal M}{(m_f=-1)}\propto J_1(\kappa b)\cdot d^{(2)}_{-1-1}(\theta_k)\approx k \theta_k/2.
\end{align}
Squaring the amplitudes and summing over all magnetic quantum numbers $m_f$ according to Eqs.(\ref{eq:xsec},\ref{eq:SpinAsym}), we obtain:
\begin{equation}
\lim_ {\theta_k\to 0}A_\Lambda^{(\overline m_\gamma=1,l_f=2)}=\frac{-1}{5+x^2},
\end{equation} 
that results in -20\% asymmetry in the vortex center, in agreement with exact results from Fig.\ref{fig:rateAs}b. We can trace the factors yielding this value of the asymmetry to the differences between Wigner $d$-functions $d^{(2)}_{21}(\theta_k)$ and $d^{(2)}_{0-1}(\theta_k)$ multiplying non-vanishing transition amplitudes in the vortex center: the latter is larger by a factor $\sqrt{3/2}$ in a small-angle limit.  Experimental observation of this difference in the excitation amplitudes can be made by analyzing normalized Rabi frequencies in ion traps, in a setup similar to Ref.\cite{Schmiegelow2016}.

Taking the expressions for CD Eq.(\ref{eq:CD}), we obtain in the paraxial limit, for example:
\begin{equation}
CD^{(\overline m_\gamma=1,l_f=2)}= \frac{4}{x^4+6 x^2+4}.
\end{equation}

Other analytic expressions in the paraxial limit for different values of the topological charge $\overline m_\gamma$ and the excited-atom OAM $l_f$ are listed in the Appendix.


\section{Summary and Discussion}			\label{sec:disc}

In this work we applied previously developed theoretical formalism \cite{Afanasev:2013kaa,Afanasev2014,Scholz2014} to analyze photoexcitation of an atom by Bessel beams with OAM and with different orientations of photon spin.
We found that both the photoexcitation rates and cross sections of twisted-photon absorption show strong dependence on the relative orientation of spin and OAM along the beam propagation direction. They can be observed by fixing the spatial structure of the beam (i.e., OAM) and flipping circular polarization with quarter-wave plates. From the parity considerations, it would be equivalent, up to an overall sign, to fixing the circular polarization and analyzing dependence on the sign of OAM projection on the beam propagation direction. Therefore our calculations predict both circular and OAM dichroism of the twisted light. Due to the factorization property of the twisted-light photoexcitation amplitudes in the first Born approximation Eq.(\ref{eq:factorized}), the plane-wave matrix elements cancel in the spin asymmetries (due to parity conservation), yielding the results independent of the internal structure of the atomic target. 

Since the rates of electric dipole transitions are proportional to the local energy flux, the corresponding CD is zero. However, position-dependent photoexcitation rates show strong dependence on the photon spin, in manifestation of spin-orbit interaction of light, {\it c.f.} \cite{BliokhAlonso10, BliokhNature15}. The corresponding rate asymmetry becomes independent of the pitch angle $\theta_k$ for moderately small angles below about 0.25 rad.  This observation provides a possible method for determination of the topological charge of optical vortex.  For the higher-multipole transitions with excitation of $l_f>1$ atomic states, the spin asymmetries are large near the optical vortex center; the asymmetries die off at distances about one wavelength from the vortex center.  For electric quadrupole transitions caused by the beams of topological charge $\overline m_\gamma$=1, the maximum spin asymmetry of the excitation rates is predicted to be -20\%, that may be observed in the experimental setup similar to Ref.\cite{Schmiegelow2016} by analyzing Rabi frequencies for individual electronic transitions in an ion trap.

We believe the present analysis of spin dependence of light absorption in atomic matter will be instrumental for quantum computing applications, optical communications, imaging, and characterization of twisted light in a broad range of wavelengths.


\begin{acknowledgments}

CEC thanks the National Science Foundation for support under Grant
PHY-1516509. Work of AA and MS was supported by Gus Weiss Endowment of The George Washington University. 
Useful discussions with K.~Bliokh, N. Litchinitser, V. Serbo, C. Schmiegelow and F. Schmidt-Kaler are gratefully acknowledged.

\end{acknowledgments}

\section{Appendix}

\subsection{Expressions for Spin Asymmetries in a Paraxial Limit}

Here we present the expressions for $CD$ and $A_\Lambda$ for small values of the pitch angle $\theta_k\to 0$, with $x\equiv k\cdot b$.

{\bf Electric Quadrupole transitions $l_f$=2}

{\it Rate Asymmetries}
\begin{align}
A_\Lambda^{(\overline m_\gamma=1,l_f=2)}&=-\frac{1}{x^2+5}, \nonumber \\
A_\Lambda^{(\overline m_\gamma=2,l_f=2)}&=-\frac{2 \left(2 x^2+9\right)}{x^4+20 x^2+18}, \\
A_\Lambda^{(\overline m_\gamma=3,l_f=2)}&=-\frac{9 \left(x^4+18 x^2+8\right)}{x^6+45 x^4+162 x^2+72}, \nonumber \\
A_\Lambda^{(\overline m_\gamma=4,l_f=2)}&=-\frac{8 \left(2 x^4+81 x^2+144\right)}{x^6+80 x^4+648 x^2+1152}. \nonumber
\end{align}
{\it Circular Dichroism}
\begin{align}
CD^{(\overline m_\gamma=1,l_f=2)}&= \frac{4}{x^4+6 x^2+4},  \nonumber \\
CD^{(\overline m_\gamma=2,l_f=2)}&= \frac{48 x^2+32}{x^6+24 x^4+84 x^2+32},  \\
CD^{(\overline m_\gamma=3,l_f=2)}&= \frac{36 \left(5 x^2+16\right)}{x^6+54 x^4+504 x^2+720}, \nonumber \\
CD^{(\overline m_\gamma=4,l_f=2)}&= \frac{64 \left(7 x^2+54\right)}{x^6+96 x^4+1744 x^2+5760}. \nonumber \\ \nonumber
\end{align}

{\bf Electric Octupole Transitions $l_f$=3}

{\it Rate Asymmetries}
\begin{align}
A_\Lambda^{(\overline m_\gamma=1,l_f=3)}&= -\frac{1}{x^2+11}\nonumber \\
A_\Lambda^{(\overline m_\gamma=2,l_f=3)}&=-\frac{4 x^2+42}{x^4+44 x^2+102}  \\
A_\Lambda^{(\overline m_\gamma=3,l_f=3)}&= -\frac{9 \left(x^4+42 x^2+80\right)}{x^6+99 x^4+918 x^2+720}\nonumber \\
A_\Lambda^{(\overline m_\gamma=4,l_f=3)}&= \nonumber -\frac{8 \left(2 x^6+189 x^4+1440 x^2+540\right)}{x^8+176 x^6+3672 x^4+11520 x^2+4320}\\ \nonumber
\end{align}
{\it Circular Dichroism}
\begin{align}
CD^{(\overline m_\gamma=1,l_f=3)}&= \frac{10}{x^4+12 x^2+10} \nonumber \\
CD^{(\overline m_\gamma=2,l_f=3)}&=  \frac{40 \left(3 x^4+8 x^2+3\right)}{x^8+48 x^6+264 x^4+320 x^2+120} \\
CD^{(\overline m_\gamma=3,l_f=3)}&= \frac{90 \left(5 x^4+64 x^2+108\right)}{x^8+108 x^6+1746 x^4+7200 x^2+9720} \nonumber \\
CD^{(\overline m_\gamma=4,l_f=3)}&=  \frac{160 \left(7 x^4+216 x^2+945\right)}{x^8+192 x^6+6304 x^4+57600 x^2+159840}\nonumber \\ \nonumber
\end{align}
\vfill



\bibliography{TwistedPhoton}

\end{document}